\documentclass[12pt]{article}

\usepackage{verbatim}
\usepackage{amssymb}
\usepackage{amsmath}
\usepackage{subfigure}
\usepackage{epsfig}
\usepackage{psfrag}
\usepackage{graphicx}
\usepackage{ulem}
\usepackage{url}
\usepackage{multirow}
\usepackage{rotating}
\usepackage{booktabs} 
\usepackage{bm} 
\usepackage{authblk}
\usepackage{tabularx}
\usepackage{color}

\DeclareGraphicsExtensions{.pdf}

\begin{document}

\title{Short Term Load Forecasts of Low Voltage Demand and the Effects of Weather}

\author[1]{Stephen Haben} 
\author[2]{Georgios Giasemidis}
\author[3]{Florian Ziel}
\author[4]{Siddharth Arora}
\affil[1]{Mathematical Institute, University of Oxford, UK}
\affil[2]{CountingLab Ltd., Reading, UK}
\affil[3]{Faculty of Economics, University Duisburg-Essen, Germany}
\affil[4]{Sa\"{i}d Business School, University of Oxford, UK}

\maketitle

\begin{abstract}
Short term load forecasts will play a key role in the implementation of smart electricity grids. They are required to optimise a wide range of potential network solutions on the low voltage (LV) grid, including integrating low carbon technologies (such as photovoltaics) and utilising battery storage devices. Despite the need for accurate LV level load forecasts, previous studies have mostly focused on forecasting at the individual household or building level using data from smart meters. In this study we provide detailed analysis of a variety of methods in terms of both point and probabilistic forecasting accuracy using data from 100 real LV feeders. Moreover, we investigate the effect of temperature (both actual and forecasts) on the accuracy of load forecasts. We present some important results on the drivers of LV forecasting accuracy that are crucial for the management of LV networks, along with an empirical comparison of forecast measures.
\end{abstract}

\section{Introduction}
\label{Sec: intro}

Increased monitoring of the electricity distribution network through advanced metering infrastructures (such as smart meters and substation monitoring) is providing better visibility and many new opportunities for managing and planning the demand on the low voltage (LV) networks. This is particularly desirable to distribution network operators (DNOs) who must prepare for the increased network stressors and distributed generation as we move to a low carbon economy. Accurate load forecasts can facilitate the management of LV networks in a number of ways including: demand side response \cite{Liu2017}, storage control \cite{Rowe2014a, Rowe2014b}, energy management systems \cite{Elbaz2015, Chitsaz2015} and integrating distributed energy resources \cite{Bennett2014}.

Load forecasting has traditionally been performed at the high voltage (HV) or system level which typically consists of the aggregated demand of hundreds of thousands or millions of consumers. The demand at such levels is much less volatile than LV demand and hence, relatively easier to predict. Load forecasting at the HV level is a very mature research area and hence, there is vast literature describing and testing a variety of techniques including, artificial neural networks (ANNs), support vector machines, ARIMA, exponential smoothing, fuzzy systems, and linear regression. For a literature review of the recent methods see \cite{Alfares2002, Taylor2008} as well as the review paper for the global energy forecasting competition (GEFCom) 2012 \cite{HONG2014357}. Many of these papers have shown strong correlations between weather effects and demand, for example, this is exhibited in \cite{Bessec2008} for a large number of European countries with the relationship dependent on whether the climate is typically warmer (as in Southern European countries) or colder such as in the UK. In \cite{Dahl2017} the authors show the link between temperature and district heating for a region in Denmark. The authors in \cite{Quilumba2015} included historical weather data in their short term load forecasts due to the strong relationships between weather and load forecasting at the system/HV level. In \cite{Dehalwar2016} the authors used weather forecasts to produce load forecasts for a large urban area in Australia. Due to the high regularity of demand at such large aggregations the mean absolute percentage errors (MAPEs) of the forecasts are typically small at around $1.5-6\%$. 

In contrast, published research on forecasting at the LV level is much more sparse due to the lack of available data. The most prominent investigations have been into short term load forecasts at the household level, applied to smart meter data. However, large quantities of smart meter data are not currently available and so much of the research has been restricted to data sets in the public domain such as from the Irish smart meter trials (\cite{Irish2012}). Due to the high volatility at the household level, it is particularly difficult to produce accurate forecasts. In fact, the authors in \cite{Haben2014} showed that due to the ``double peak'' error for spiky data sets it is difficult to measure the accuracy of household level point forecasts using traditional pointwise errors. As with the HV level, similar methods have been applied to the household level, including ANNs \cite{Sousa2009}, ARIMAs \cite{Veit2014}, wavelets \cite{Chitsaz2015}, Kalman filters \cite{Ghofrani2011}, and Holt-Winters exponential smoothing \cite{Yu2017}. The errors in these methods are much larger than the HV level, with MAPEs ranging from $7\%$ up to $85\%$ \cite{Sousa2009, Veit2014}. A link between weather and household demand has been observed and historical weather data has been used to generate forecasts \cite{Yu2017, Liu2017}. Some of the strongest correlations observed have been in temperature and illumination \cite{Hobby2011}. 

The literature on LV level forecast research is sparse compared to household/smart meter level forecasting. LV distribution feeders are relatively volatile compared to HV level demand since they consist of low aggregations of customers (typically less than $150$ in the UK) \cite{Giasemidis2017}. The main forecasting research has been presented in \cite{Bennett2014, Bennett2014b} where the authors apply both ARIMAX and ANN methods to a single LV transformer (consisting of 128 customers) to forecast total energy and peak demand. They achieve MAPEs of between $6-12\%$. They also included historical weather data in the methods. At slightly higher voltage substations, the authors in \cite{Ding2016} and \cite{Alberg2017} apply ANN and ARIMA methods to MV/LV level data to achieve MAPEs of $11\%$ and $13-16\%$ respectively. The majority of load forecasts at the LV substation level are in fact applied to aggregates of smart meter data \cite{Humeau2013, Sevlian2014, Zufferey2017}. In \cite{Humeau2013, Sevlian2014} the authors consider a variety of methods and show a strong scaling law relationship between the MAPEs and the aggregation size. In \cite{Sevlian2014} the authors consider the aggregation level of data from 1 to 100M smart meters. The relationship between relative accuracy and aggregation is verified in \cite{Zufferey2017} where the authors consider aggregations of 1, 10, 100, 1000, 10000 smart meters using ANNs applied to data from 40,000 customers in Basel, Switzerland. MAPEs vary considerably from as low as $2\%$ \cite{Alzate2013} up to $30\%$ \cite{Hayes2015}. The relationship between temperature and load has had mixed results at the LV level (based on aggregate data rather than LV level data). In \cite{Hayes2015} the authors successfully apply weather data with ANN and ARMA methods to 4 different levels of aggregations (1, 10, 100, 1000 smart meters) from both the Irish and a Danish smart meter set. In \cite{Valgaev2016}, the authors consider aggregates of 5, 10 and 100 smart meters from the Irish data set. They do not utilise weather data as they argue that the weather data is not scalable. In \cite{Dangha2017} the authors consider two main data sets consisting of 40 time series and apply ARIMA, Holt-Winters, ANN, and generalised additive models. Although they show there is a correlation between weather and load in the data sets, they suggest that there is not much affect of temperature on load.

Due to the volatility of LV level demand, probabilistic load forecasts are a natural choice to provide a detailed description of the underlying uncertainty. Most recently there has been an increased interest in probabilistic load forecasting, accelerated by the recent global energy forecasting competition 2014 \cite{Hong2016}. See \cite{Hong2016b} for an up to date review of the current state of the art and major challenges in probabilistic load forecasting.  There have been a few publications where the authors have considered load forecasting of individual smart meter data or on aggregations of smart meter data. In \cite{Arora2016} the authors applied kernel density estimation methods to the Irish smart meter data and compared the errors between the forecasts of domestic and non-domestic customers over a horizon of up to a week ahead. In \cite{Gerossier2017}, probabilistic forecasts of smart meter data from 226 Portuguese households were considered using quantile regression with a generalised additive model (GAM). In \cite{Taieb2016} the authors also use a GAM considering both individual and aggregations of the Irish smart meter data up to 1000 smart meters. They found at large aggregations that a normal distribution was a sufficient model for the demand, thus demonstrating the law of large numbers and the contrast to the LV level demand. The same authors also considered various aggregation levels of UK smart meter data in \cite{Taieb2017} using copulas to develop the joint distributions at different levels of aggregation to ensure coherent probabilistic forecasts. Finally, in \cite{Li2017} day ahead quantile forecasts are created using Laplace distributions and non-parametric methods. Although probabilistic forecasts are more informative for DNOs to make better informed management decisions, they are also much more computationally expensive. There are costs associated with generating the probabilistic forecasts and selecting the appropriate models. In \cite{Xie2016} a comparison is made between model selection in point versus probabilistic scoring functions. They compare MAPE against the pinball function and show that similarly accurate models are produced in both. This could have important implications for reducing the computational cost of model/parameter selections. In the research presented here we consider point and probabilistic forecasts and compare both point scoring functions (such as MAPE) to probabilistic scoring functions (continuous ranked probability score). We show that it may be sufficient to evaluate accurate point forecasts to produce corresponding accurate probabilistic forecasts.

The aim of this paper is to build upon the current research and present detailed point and probabilistic load forecasts and rigorous analysis for a large number of real LV level substation feeders based in a town in the South of England. We generate and compare a large number of methods and benchmarks to better understand the drivers of accurate LV level forecasts and identify important features. We also consider the relationship between the size of the feeder and the accuracy of the forecasts, confirming the power law relationship as found in the literature \cite{Sevlian2014}. In contrast to aggregated data, LV networks consist of different proportions of a variety of customers (domestic, small-to-medium enterprises (SMEs) and commercial) as well as other, unmonitored street furniture such as street lighting and traffic lights \cite{Sun2016}. Hence, there is a limit to the representativeness of the aggregated results. Another contribution of this paper is with respect to the investigation of temperature effects. Firstly, we consider models with and without temperature. This provides an opportunity to investigate whether the relationship between load and temperature is causal or simply a correlation effect, which has not been considered in much detail in the literature. Secondly, in this study we have access to temperature forecasts in the region of interest. In much of the load forecasting literature, weather forecast data is not always available and historical weather is used instead. This limits the feasibility and conclusions since such forecasts are not possible in practice. Here we will be able to compare the difference between using historical and forecast temperature in our load forecasts. 

The rest of the paper is organised as follows: In section \ref{Sec:DataAnal} we analyse the data that we use in this study. In section \ref{Methods} we describe the methods we will be using as well as the scoring functions we use to measure the accuracies of our forecasts. In section \ref{Sec:Results} we describe the main results and in the final section there is a discussion including potential future work.

\section{Data Analysis}
\label{Sec:DataAnal}

In this section we review and analyse the data that will be used to create and evaluate our methods. 

\subsection{Load Data}

The load data for the 100 feeders begins on $20^{th}$ March $2014$ up to the $22^{nd}$ November 2015, a total of $612$ days, based in Bracknell, a town in the south of England. The data was collected as part of a low carbon network fund project called the \textit{Thames Valley Vision}\footnote{see \url{http://www.thamesvalleyvision.co.uk/} for more details.}. The data from each feeder is sampled hourly. The feeders consist of a range of magnitudes with the average daily demand of approximately $602$kWh and a maximum and minimum daily demand of around $1871$kWh and $107$kWh respectively. Of these feeders 83 of the 100 are connected purely to residential consumers with an average of 45 households, a maximum of 109 customers and a minimum of 8. A further 7 have no connectivity information. For the trial presented in this paper we define a test set, consisting of the dates $1^{st}$ October 2015 to $22^{nd}$ November 2015 inclusive. The remainder of the data is used for training. 

\begin{figure}
\begin{center}
\includegraphics[width=10cm]{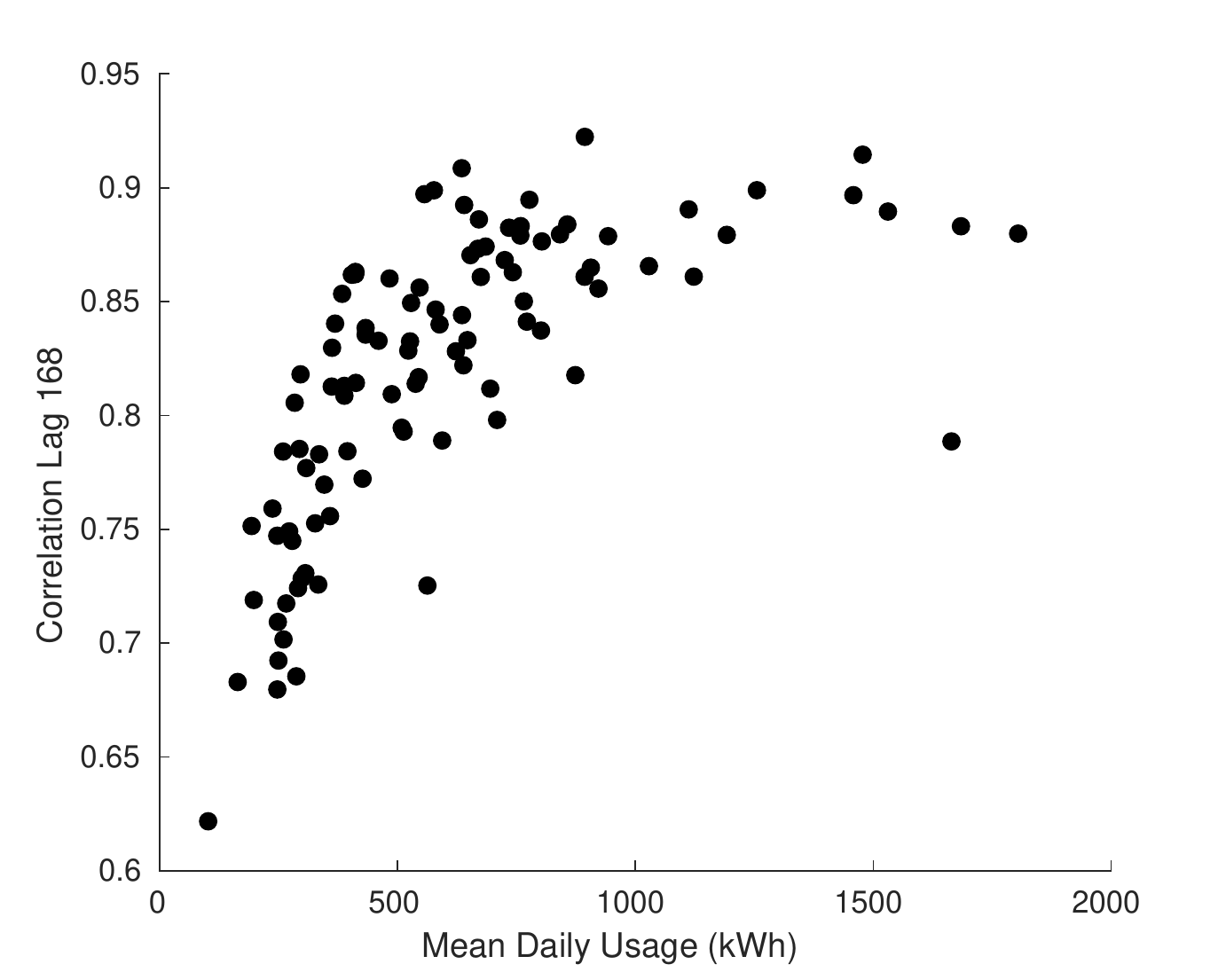}
\caption{\label{AutocorrelationLag168vMeanDemand} Autocorrelation at lag 168 for all 100 feeders against the mean daily demand.}
\end{center}
\end{figure}

The data contains strong weekly and daily periodicity. Figure \ref{AutocorrelationLag168vMeanDemand} shows the relationship between the autocorrelation at lag 168 (i.e. a week) and the mean daily feeder demand. The plot highlights that all feeders have some degree of weekly regularity with the larger feeders tending to have much stronger autocorrelation than smaller feeders, and hence, compared to smaller feeders one might expect the larger feeders to have lower forecast errors.

There is also a strong seasonal effect in the data, however note, that due to the lack of air conditioning in the UK there is no increase in demand for warmer periods in contrast to other data sets which are commonly shown in the literature \cite{Hong2016}. 

\subsection{Weather Forecasts}

Weather variables, especially temperature, and those related to temperature such as wind chill, are often included within load forecasting models \cite{Rahman1990, Bessec2008}. Typically those are for high voltage load forecasts and hence represent a large numbers of customers demand. However, it is not obvious that weather variables play an important role in load forecasts at the LV level. In this report we consider temperature data (in degrees), collected from the Farnborough weather station, the closest weather station to Bracknell at just under 10 miles (approx 16 Km). In our forecasts we will utilise both the historical and forecast temperature data.  

Although temperature has a strong correlation with demand at the high voltage, for practical load forecasts, only temperature forecasts can be used. We first consider the accuracy of such forecasts before considering the correlation of temperature with the LV load. The forecasts all begin at $7$am on each day and forecast each hour for the next four days. In other words, the one hour ahead forecasts are all for the period $8$am, the two hour ahead forecasts are all at $9$am, etc. Thus, the temperature forecast accuracy is not simply determined by the forecast horizon (where usually greater forecast horizons correspond to more inaccurate forecasts), but also the volatility of the time period being estimated. Indeed the accuracy (as measured by the mean absolute percentage error) does reduce as a function of horizon, as shown in Table \ref{TempFor}, for the full data set. Forecasting temperature four days ahead (i.e. between 73 and 96 hours ahead) has reduced the accuracy by around $100\%$ compared to up to one day ahead (between 1 and 24 hours ahead). For the test set the temperature forecast accuracy is improved compared to the training set (see Table \ref{MAPErrorsSSPerDay2} in Section \ref{sec:Forecast Weather}).

\begin {table}[h]
\begin{center}
  \begin{tabular}{ |c|c|c|c|c| }
\hline
Forecast Horizon & 1 Day & 2 Day & 3 Day & 4 Day \\
\hline
MAPE   			& 11.85  	&  15.60	& 20.21 & 23.80 \\
\hline
  \end{tabular}
\end{center}
\caption{\label{TempFor} MAPE for the temperature forecasts for different horizon periods (over the period $20^{th}$ March 2014 to $28^{th}$ Nov 2015).} 
\end{table}

In our analysis we will focus on the relationship between the load forecasts using forecast temperature inputs (ex-ante forecasts) rather than the historical temperature inputs (ex-post forecasts) \cite{Hong2014}. However, although the forecast and actual temperature values are very strongly correlated (with a pairwise linear correlation coefficient greater than 0.95) we will include results for both ex-ante and ex-post versions. We do this for two reasons. Firstly, much of the literature presents ex-post forecasts (i.e. those which use the historical temperature inputs) and hence we can compare our results with other research \cite{Hong2014}. Secondly, it provides further evidence concerning the causal link between weather and load at the LV level. 

\begin{figure}
\begin{center}
\includegraphics[width=12cm]{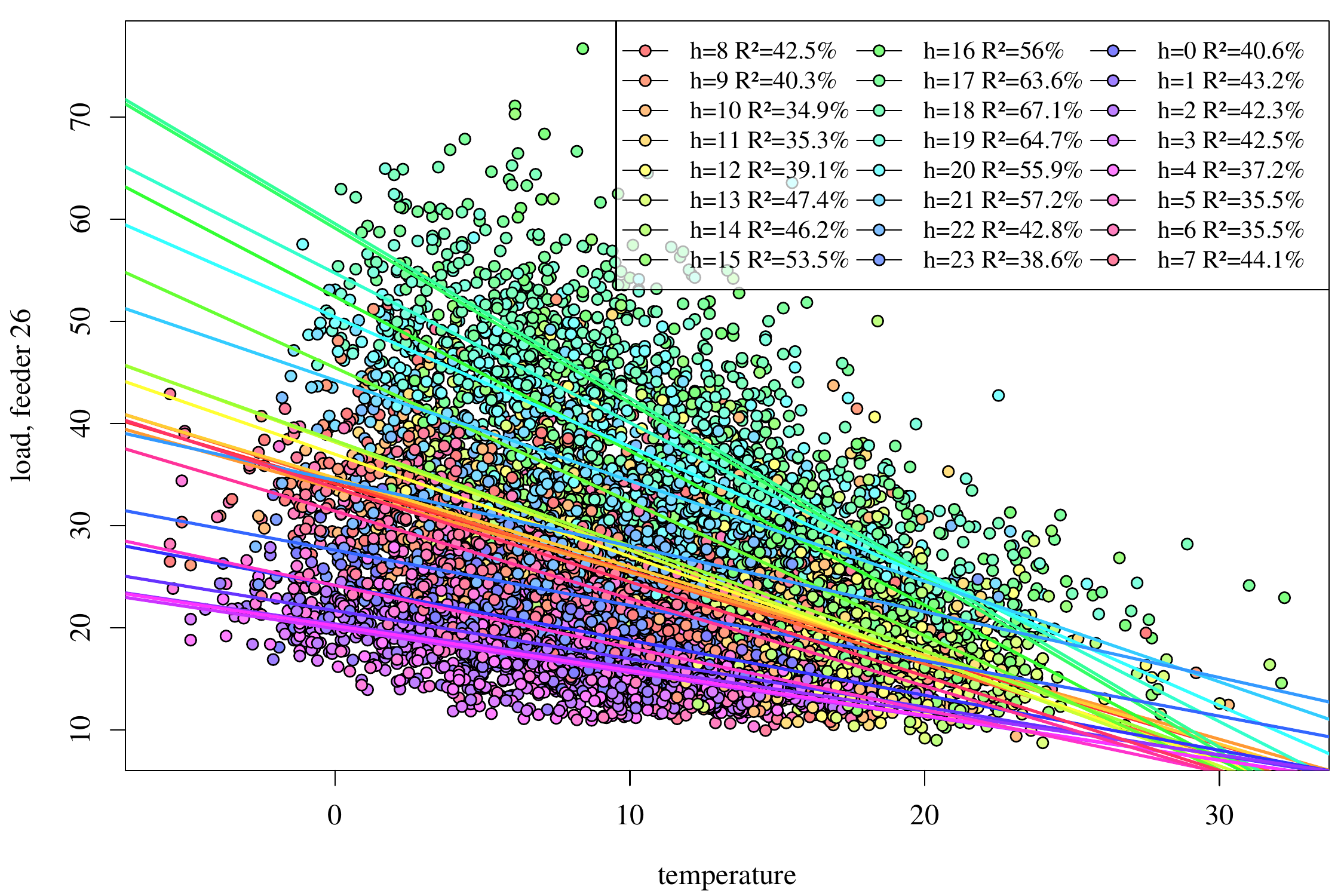}
\caption{\label{load_temp_24_feeder1026} Load and corresponding temperature for feeder 26. The best linear fits and $R^2$ values are shown and colour coded according to different hourly periods of the day.}
\end{center}
\end{figure}

As with high voltage load, there are correlations, usually negative, between temperature and the LV loads from our feeders. Comparing the full time series the average correlation between the load and the temperature is $-0.14$ with the strongest correlation of $-0.63$. However, different time periods of the day are more strongly correlated with temperature than others. If we now split each of the 100 time series into 24 time series (one for each hour of the day) the average correlation is $-0.47$ with the strongest correlation between any time period across all feeders of $-0.76$. A strong correlation relationship between the demand and temperature is shown in Figure \ref{load_temp_24_feeder1026} for a particular feeder in our trial. The values are broken down according to hourly periods of the day and include a basic linear fit and $R^{2}$ values.  

This section has presented some basic analysis of the load data and corresponding temperature data. We have highlighted some important features which should be utilised in the load forecasts. Daily, weekly and annual periodicities have strong relationships to load demand and there are different behaviours in demands for different periods of the day. Further we have seen there is a wide variety of feeders (in terms of numbers of connected customers and magnitudes of demands) with different correlations with the temperature. We will use this information to construct our forecasts as well as appropriate benchmarks. 

\section{Methods} \label{Methods}

In this section we present a wide variety of methods which will be used to generate the forecasts. A number of benchmarks are also included for comparison as well as to highlight the importance of various inputs/predictors. The chosen methods are motivated by the data analysis presented in Section \ref{Sec:DataAnal}, and the success of load forecasting methods used in the load forecasting literature, including the global energy forecasting competition (GEFCom) 2014 \cite{Hong2016}. 

For this section, without loss of generality, we define $L_t$, $t=1, ...,N=D\cdot H$ to be the hourly time series for the load for a particular feeder, where $H=24$ and $D=612$ is the number of days in the training and test data set combined. The initial time step $t=1$ defines the start of the data set on the $20^{th}$ March 2014 for the hourly period $12AM$ to $1$AM. Also, without loss of generality, we also define $t_h$ to denote the end of the historical data (with $t_h+1$ therefore the start of the testing period) which determines the maximum available training data for the methods. Specifics of each methods will be described in their corresponding sections.

\subsection{Kernel Density Estimation} \label{Sec:KDE}

The first method we consider are those based on kernel density estimation (KDE) techniques. These have been successful at generating probabilistic forecasts for individual smart meter data as well as in the GEFCom2014 competition on higher voltage demand \cite{Arora2016, Haben2016b}. The basic aims of the method is to generate a conditional distribution at time $t$, $f(L_t|X)$, conditional on historical demand and other factors, $X$, such as period of the week or weather variables. The main challenge of such methods is the computational expense of evaluating the correct parameters, in particular the bandwidths. 

In this section we presume the parameters are trained on the historical data $[t_1, t_2]$, with $1\leq t_1 < t_2 \leq t_h$, $t_1 \mod H=1$ and $t_2 \mod H=0$, in other words the period of time encapsulates full days from the historical data. There are a number of available choices for the kernels used in the methods. Here, we simply use the Gaussian kernel, which have been successfully used in other implementations of load forecasting \cite{Haben2016b}, and secondly, evidence in the literature suggests that the choice of kernel has minimal effect on the accuracy of the forecasts \cite{Jeon2012}. 

The first method uses a kernel density estimate (KDE) to estimate the probability density function for each period $w= t \mod 24$ of the day, defined by the function
\begin{equation}
f(\hat{L}_t)=\frac{1}{(t_2-t_1+1)h_L}\sum_{i\in I_w} w_i K \left( \frac{L_t-L_i}{h_L} \right), 
\label{KDEWlambda}
\end{equation}
where $I_w=\{i \in [t_1, t_2]| i \mod 7H = w\}$ is the index set denoting time periods from the training set from the same period of the week as the time period $t$, and $w_i$ is the weighting of historical observations $i$. We consider two types of KDE forecasts. The first type, denoted \textbf{KDE-W}, considers all observations having equal weight, $w_i=1$ for all $i$. This forecast has one parameter, the bandwidth, $h_L$, for the load kernel that requires optimisation.  The second type, referred to as \textbf{KDE-W$\mathbf{\lambda}$}, favours observations around the same period of the year as time $t$, i.e. $w_i = \frac{\lambda^{\alpha(i)}}{\sum_{i\in I_w}\lambda^{\alpha(i)}}$,  with the decay exponent $\alpha(i)$ defined by
\begin{eqnarray}
\alpha(i) &=& \min \left (|\mathcal{W}(t) - \mathcal{W}(i)|, 52 - |\mathcal{W}(t) - \mathcal{W}(i)|\right ),
\label{expoKDEW}
\end{eqnarray}
where $\mathcal{W}(i) \in \{ 1, 2, \ldots, 52\}$ is the week of the year corresponding to the load data, $L_i$. Note, the first Monday of the year is defined as $\mathcal{W}(i) = 1$. Equation \eqref{expoKDEW} is simply a periodic absolute value function with annual period, whose minimum values occur annually on the same week as the estimated day. The exponent is more relevant when there are several years of historical data. KDE-W$\lambda$ has two parameters to optimise, $h_L$, the bandwidth for the load kernel and $\lambda \in (0, 1]$.

Additionally, we consider kernel density estimate forecasts conditioned on independent variables $y, z$ (CKD), such as the week-period, or weather variables e.g. the temperature or both. This is represented as 
\begin{equation}
f(\hat{L}_t|y, z)=\sum_{i \in [t_1, t_2]}
\frac{ K((y_i-y)/h_y) K((z_i-z)/h_z)}{\sum_{i=1}^{n} K((y_i-y)/h_y) K((z_i-z)/h_z)} 
K \left( \frac{L_t-L_i}{h_L} \right)
\label{eq:CKD}
\end{equation}
where $h_y, h_z$ are the bandwidths of the independent variables $y, z$ respectively. If there is only one independent variable $y$, then one ignores the kernel of $z$ in \eqref{eq:CKD}. CKD methods consider the whole time-series of historical observations at all time-periods.

We produce three CKD forecasts, one conditioned on the week period (\textbf{CKD-W}), a second conditioned on the week period and the actual temperature readings (\textbf{CKD-WTa}), and a third forecast conditioned on the both the week period and the forecast temperature (\textbf{CKD-WTf}). For CKD-W, $y_i = i \mod 7H$ is the week period of time interval $i$. CKD-W weighs observations towards similar times of the week as the forecast time-period.

The CKD methods have not been implemented with a decay parameter due to the increased computational cost. Any extra parameter increases the dimension of the parameter space and hence the computational cost for optimisation. The bandwidths and/or $\lambda$ parameters of each method are found via validation. The validation period is selected to be two weeks prior to the test-period. For the KDE forecasts, all available observations before the two-week validation period are considered for training. As CKD methods are computationally more expensive, we restricted the training period to a year prior to the validation period. Finding the optimal parameters is a non-linear optimisation problem. For KDE-W, the function \textit{fminbnd}\footnote{\url{https://uk.mathworks.com/help/matlab/ref/fminbnd.html}} MATLAB's built-in optimisation algorithm, was used. For KDE-W$\lambda$ and all the CKD forecasts, the \textit{fminsearchbnd}\footnote{\url{https://uk.mathworks.com/matlabcentral/fileexchange/8277-fminsearchbnd--fminsearchcon}} optimisation package was used instead.

The load and temperature variables are normalised to $[0, 1]$ for each feeder, to accelerate the optimisation procedure, as the parameter space is restricted to the interval $[0, 1]$. When the optimisation is complete and a forecast is produced, the normalised forecast is rescaled. To assess the effect of normalisation in the forecasts, we also experimented with KDE-W and KDE-W$\lambda$ forecasts without normalising the load. We compared the two forecasts and their errors. The forecasts were almost identical, with marginal error differences, evenly distributed around zero. However, the optimisation with normalised variables requires, on average, 6 less iterations than the optimisation with actual readings. For this reason, we decide to use normalised load and temperature variables for the parameter optimisation of all methods.

An advantage of the kernel-density methods is that the entire distribution is found simultaneously whereas quantile regression methods only find individual quantiles. The disadvantage is the computational costs, especially as more conditional variables are introduced. Various methods, such as clustering the time series, must be employed to reduce the costs \cite{Arora2016}. We consider the KDE based
methods to generate probabilistic forecasts. We use the median of the forecast distribution as a point forecast.

\subsection{Simple Seasonal Linear Regression}
\label{sec:ST}

The method is based on an update of the simple seasonal model presented in \cite{Haben2016b}. For this method rather than construct a full probability density function for the load distribution we instead develop models for a number of predefined quaniltes $\tau \in (0, 1)$. Here, we assume a linear regression model for each quantile, treating each period of the week as a separate time series of the form
\begin{eqnarray}
\hat{L}^{\tau}_t &=& \sum_{k=1}^{H}\mathcal{D}_{k}(t)\left(a^{\tau}_k+b^{\tau}_k \eta(t)+\sum_{p=1}^P (c^{\tau}_k)^p\sin \left( \frac{2\pi p \eta(t)}{365} \right) +(d^{\tau}_k)^p\cos \left( \frac{2\pi p \eta(t)}{365} \right) \right) \nonumber \\
& & + \sum_{l=1}^{7H}f^{\tau}_l\mathcal{W}_l(t).
\label{eq:oldLoad}
\end{eqnarray}
Here $\eta(k)=\left\lfloor \frac{t}{H}\right\rfloor+1$, is the day of the trial (with day 1 as $20^{th}$ March 2014). There are two dummy variables identifying the period of the day, $\mathcal{D}_{k}(t)$, and the period of the week, $\mathcal{W}_l(t)$, defined by
$$\mathcal{D}_j(t) = \begin{cases}
       1, & t\mod H = j, \\
	  0, & \text{otherwise},
           \end{cases}
$$
and
$$\mathcal{W}_j(t) = \begin{cases}
       1, & t\mod 7H = j, \\
	  0, & \text{otherwise},
           \end{cases}
$$
respectively. 

There are essentially $168$ models representing each period of the week each with average, linear trend, and annual seasonality terms. The $a_k$ terms represent the average demand for that half hourly period (which is augmented based on the day of the week by $f_l$), a linear trend term $b_k$, and annual seasonality terms defined by $c_k$ and $d_k$. The parameters for each hour and each quantile are found by a quantile regression over the historical data using the pinball function \cite{Koenker1978}. For the point forecast as with the KDE methods we considered the median quantile and also a least squares estimate. In all cases the median outperforms the least squares estimate and hence our point forecast estimates will be derived from the median quantile. The methods are trained on the entire available historical information using the latest data for the rolling forecasts at the start of each new day.

We considered a number of variants of the model presented here including using different numbers of seasonal terms ($P=2, 3$), with and without trend (i.e. $b_k\equiv 0, \forall k$), and also using only weekend dummy variables instead of dummy variables for all days of the week. We found that the best methods used a linear trend, three seasonal terms ($P=3$), and used the day of the dummy variable as in equation (\ref{eq:oldLoad}). However, we also found that the model without trend also performed reasonably well, so in our analysis we will consider both methods. We will denote the seasonal method with trend as \textbf{ST} and without trend as \textbf{SnT}. 

Including temperature effects in the model is straightforward and only requires adding a polynomial to the full equation (in our case up to only cubic order). Depending on the horizon (one, two, three, or four days ahead) four different models are calibrated.

\subsection{Autoregressive Methods}
\label{sec:autoregpoint}
The methods in the previous sections only take into account the potential autocorrelation within the time series in a limited way. For CKD-W the correlation is focused around the same periods of the week and the ST method contains no correlation with other time periods. The methods presented here will include stronger autocorrelation effects within the models. We first describe two methods for creating point forecasts. 

The models are all based on a regression of a residual time series $r_t=L_t-\mu_t$, for some \textit{mean} profile $\mu_t$ which we define later. This residual regression can be written,
\begin{equation}
r_t = \sum_{k=1}^{p_{\max}} \phi_k r_{t-k} + \epsilon_t
\label{MeanEQNRes}
\end{equation}
for some Gaussian error $\epsilon_t$. The main tunable parameter in the model is the optimal autoregressive order $p=p_{\max}$ which is chosen by minimising the Akaike information criterion (AIC) over $p \in \{0, \ldots, p_{\max}\}$, then the coefficients $\phi_1, \ldots, \phi_{\max}$ can then be easily determined by the Burg method. The methods are trained over a years worth of historical data prior to the start of the test period on the $1^{st}$ October 2015. 

For the mean profile $\mu_t$ we consider two models. The first estimates a simple weekly average and can be written
\begin{equation}
 \mu_t = \sum_{j=1}^{7H} \beta_{j} \mathcal{W}_j(t).
\label{MeanEq}
\end{equation}
where 
$\mathcal{W}_j(t)$ is the period of the week dummy variable as defined in Section \ref{sec:ST}. The weekly mean parameters are $\beta_{j}$ and the mean $\mu_{k}$ equation (\ref{MeanEq}) is estimated by ordinary least squares (OLS) over the initial prior year of historical loads. We denote the model \textbf{ARWD}.

The second model updates the simple weekly average to include an annual seasonality. 
\begin{equation}
 \mu_t = \sum_{j=1}^{7H} \beta_{j} \mathcal{W}_j(t) + \sum_{k=1}^K \alpha_{1,k} \sin( 2\pi t k/ A ) + \alpha_{2,k} \cos( 2\pi t k/ A )
\label{MeanEq2}
\end{equation}
with parameters $\beta_{j}$ and $\alpha_{j,k}$, and $A=365H$ as annual seasonality. The annual seasonality is modelled by a Fourier approximation of order $K$ (we choose only $K=2$). The dummy variable $\mathcal{W}_j(t)$ is as before. As with the \textbf{ARWD} model the $\mu_{k}$ is estimated by a OLS regression fit to the training data. We denote the model by \textbf{ARWDY}.

As with the ST methods it is trivial to include weather effects by adding linear terms to the mean equations. We note that seperate models are used depending on whether the forecasts are one, two, three or four days ahead.

The methods can be updated to generate probabilistic forecast methods. The methods are not quite as straightforward as the point forecasts. We present a related but alternative autoregressive probabilistic forecast technique for the ARWDY case (which generalises trivially for the ARWD case). 

The following method creates a distribution by modelling the variance of the model from the residuals. As before an autoregression is performed on the residual equation (\ref{MeanEQNRes}) but now the mean load is modelled by the following equation. 
\begin{equation}
 \mu_t = \sum_{j=1}^{7H} \beta_{j} \mathbb{W}_j(t) +
 \sum_{k=1}^K \gamma_{k}\mathbb{D}_j(t) +\alpha_{1,k}\mathbb{D}_ j(t)\sin( 2\pi t k/ A ) + \alpha_{2,k} \mathbb{D}_j(t) \cos( 2\pi t k/ A ), 
 \label{Probmeaneqn}
\end{equation}
where $A=365H$. The annual seasonality is modelled by a Fourier approximation of order $K=2$. This time a different (but equivalent) basis function is used with dummy variables for different horizon periods modelled as 
$$\mathbb{W}_j(t) = \begin{cases}
            1, &  \text{t mod 7S} \leq j \\
	    0, & \text{otherwise}
           \end{cases}
         \ \  \text{ and } \ \ 
           \mathbb{D}_j(t) = \begin{cases}
            1, & \text{t mod S} \leq j \\
	    0, & \text{otherwise}
           \end{cases}.
$$
The mean equation (\ref{Probmeaneqn}) is solved using a lasso method tuned to ensure that the Hannan-Quinn information criterion (HQC) is minimised. The past year is used in the training. The equation (\ref{MeanEQNRes}) is then estimated using the discovered residuals by solving the Yule-Walker equations. As before, the autoregressive order is chosen by minimising the Akaike information criterion. 
To create the confidence bounds we model the error term (in equation (\ref{MeanEQNRes})) as 
\begin{equation}
\epsilon_{t} = \sigma_{t} Z_t,
\label{eqn:eps}
\end{equation}
where $\sigma_t$ is the conditional standard deviation of $\epsilon_t$ and $(Z_t)_{t_\in \mathbb{Z}}$ is an independent and identically distributed (iid) random variable with $\mathbb{E}(Z_t)=0$ with $\mathbb{V}ar(Z_t)=1$. We then assume a similar relationship for the standard deviation as we do for the mean, 
\begin{equation}
\sigma_t = \sum_{j=1}^{7H} \tilde{\beta}_{j} \mathbb{W}_j(t) +
 \sum_{k=1}^K \tilde{\gamma}_{k}\mathbb{D}_j(t) +\tilde{\alpha}_{1,k}\mathbb{D}_j(t)\sin( 2\pi t k/ A ) + \tilde{\alpha}_{2,k} \mathbb{D}_j(t) \cos( 2\pi t k/ A), 
 \label{Probstdeqn}
\end{equation}
Although we do not know the values of the variance we can model a scaled version of $\sigma_t$ by using the model in equation (\ref{Probstdeqn}) to fit (using a lasso method to minimise the HQC) the model for $|\epsilon_t|$. Since $\mathbb{E}(|\epsilon_t|)=\sigma_t\mathbb{E}(|Z_t|))$ this means we are actually estimating $C\sigma_t$ for some constant $C$. We estimate the constant by considering the residuals $\epsilon_t/C\sigma_t$ which, given (\ref{eqn:eps}), should therefore behave like a scaled version of $Z_t$. Considering the variance of these residuals and noting that $Z_t$ has variance one, we use this to estimate the constant and hence find $\sigma_t$. We now calculate the standardised residuals $\epsilon_t/\sigma_t$ to put an empirical distribution on $Z_t$ and estimate the quantiles (which once scaled by $\sigma_t$ give the quantiles on $\epsilon_t$). Note, if $Z_t$ follows a standard normal distribution then C would be $\sqrt{2/pi} = 0.798.$ For our data we found this was usually smaller, between 0.5 and 0.75. An advantage of this method is the quick computational speed. Computing 99-percentiles for 4-days ahead takes about 2 seconds per feeder.

\subsection{HWT Exponential Smoothing Method}
\label{sec:HWTmethods}
We implement the Holt-Winters-Taylor (HWT) exponential smoothing method \cite{Taylor2003} to model the intraday and intraweek seasonality in the feeder load. This method is represented as:
\begin{eqnarray}
L_{t} &=& l_{t-1} + d_{t-s_1}+w_{t-s_2}+\phi e_{t-1} + \epsilon_{t}  \nonumber\\
e_{t} &=& L_{t} - (l_{t-1} + d_{t-s_1}+w_{t-s_2}) \nonumber \\
l_{t} &=& l_{t-1} + \lambda e_{t} \nonumber \\
d_{t} &=& d_{t-1} + \delta e_{t} \nonumber \\
w_{t} &=& w_{t-1} + \omega e_{t} 
\label{eq:HWT}
\end{eqnarray}
where $L_{t}$ denotes the load observed at time $t$, $l_{t}$ denotes the level, $\epsilon_{t} \sim$ IID(0,$\sigma^2$), $s_1$ = 24, $s_2$ = 168, and $d_{t}$ and $w_{t}$ correspond to the intraday and intraweek seasonal indexes, respectively. This model requires the estimation of three smoothing parameters $\lambda$, $\delta$ and $\omega$ for the level and two seasonal indexes, along with a parameter $\phi$ to adjust for first order auto-correlation in the error (denoted by $e_{t}$). We estimate model parameters by minimizing the sum of one-step ahead sum of squared errors using the training data. The model parameters were estimated separately for each feeder. To generate out-of-sample probability density estimates for the load we used Monte Carlo simulation, by constructing an ensemble of $1000$ scenarios. The median of density forecasts was issued as a point forecast for each observation in the test data. The interpolated load values were ignored during the process of both model estimation and evaluation.

\subsection{Benchmark Methods}
\label{sec: Benchmarks}

As a comparison to the methods presented we implemented several simple benchmark to model load for each feeder. The details of these benchmarks are presented below:
 
\subsubsection{Seasonal random walk for same period of the day (\textbf{LD})}
To forecast load for a given period, we issue the load observed on the same period from the last day as a forecast. We use:
\begin{equation}
\hat{L}_{t+k} = L_{t+k-s_1}  
\end{equation}
where $\hat{L}_{t+k}$ is the $k$-step ahead prediction, $t$ is the forecast origin, and $s_1=24$ is the intraday cycle length.

\subsubsection{Seasonal random walk for same period of the week (\textbf{LW})}
We issue the load observed on the same period from the last week as a forecast. We use:
\begin{equation}
\hat{L}_{t+k} = L_{t+k-s_2}  
\end{equation}
where $s_2=168$ is the intraweek cycle length.

\subsubsection{Seasonal random walk for same period of the year (\textbf{LY})}
We issue the load observed on the same period from the last year as a forecast. We use:
\begin{equation}
\hat{L}_{t+k} = L_{t+k-s_3}  
\end{equation}
where $s_3=52 \times s_2$ is the intrayear cycle length.

\subsubsection{Seasonal moving average using a p week period (\textbf{SMA-pW})}
We issue the mean of load observed on the same period during the last four weeks as a forecast. We use:
\begin{equation}
\hat{L}_{t+k} = \frac{1}{p}\sum_{i=1}^{p}L_{t+k-i \times s_2}  
\end{equation}
We considered a variety of different p values but the best were for $p=4,5$ weeks. 

\subsubsection{Seasonal moving average using an optimal window period (\textbf{SMA})}
We issue the mean of load observed on the same period during the last $p$ weeks as a forecast. We use:
\begin{equation}
\hat{L}_{t+k} = \frac{1}{p}\sum_{i=1}^{p}L_{t+k-i \times s_2}  
\end{equation}
The optimal value of $p$ was estimated separately for each feeder, which was based on the minimization of one-step ahead sum of squared errors. The last one month of the training data was used as a cross-validation hold-out-sample. Some of these benchmarks were used in a case study for modeling French load \cite{Arora2017}.

\subsubsection{Empirical Estimate}

The other benchmark methods only provide point estimates and hence cannot provide quality comparisons to probabilistic forecasts. For each period of the week we define an empirical distribution function using all the load data from the same time period over the final year of the historical data. We then use this empirical distribution to define the desired quantiles. The median quantile is used as a point estimate. The estimate is fixed over the entire test period. We refer to this method as the \textbf{Empirical} forecast. Note that we could have included more historical data in the construction of the empirical distribution but by restricting it to the last year we do not produce seasonal biases.

\subsection{Error Measures}
\label{Sec: ErrorMeasures}

To evaluate our methods we consider a variety of forecast measures which are common to the forecasting and in particular, load forecasting literature. Suppose we have actual loads $\mathbf{a}=(a_1, a_2, \ldots, a_n)^T$ for time periods $\hat{t}_1, \hat{t}_2, \ldots, \hat{t}_n $ and forecasts $\mathbf{f}=(f_1, f_2, \ldots, f_n)^T$ for the same time period. The most commonly used score for evaluating the accuracy of point load forecasts is the mean absolute percentage error (MAPE)
\begin{equation}
MAPE(\mathbf{a}, \mathbf{f})= \frac{100}{n}\sum_{k=1}^n\frac{|a_k-f_k|}{|a_k|}.
\label{MAPE}
\end{equation}
Another common error measure which we will consider is the mean absolute error (MAE) which is defined as
\begin{equation}
MAE(\mathbf{a}, \mathbf{f})= \frac{1}{n}\sum_{k=1}^n |a_k-f_k|.
\label{MAE}
\end{equation}

The MAPE and MAE are only applicable to the point forecasts. To estimate the accuracy of the probabilistic forecasts, given by a cumulative distribution $F(z_k)$ at time $\hat{t}_k$, we use the continuous ranked probability score (CRPS) which quantifies both calibration and sharpness of the forecasts \cite{Gneiting2007} and is defined by
\begin{equation}
CRPS(\mathbf{a}, \mathbf{f})= \frac{1}{n}\sum_{k=1}^n\int_{-\infty}^\infty (F(z_k) -\mathbf{1}(z_k-a_k))^2 dz_k,
\label{CRPS}
\end{equation}
where $\mathbf{1}$ is the Heaviside step function. 
The MAE and CRPS are scale dependent and therefore we cannot compare feeders of different fundamental sizes. We therefore normalise the score by dividing the MAE and CRPS by the average hourly load of the feeder over the last year of training data. We also multiply the errors by 100 so that they may be referred to as percentages. We will refer to these as the relative MAE (RMAE) and relative CRPS (RCRPS) respectively. The CRPS reduces to the MAE in the case of a point forecast and hence we expect them to be strongly related \cite{Gneiting2007}. The MAPE scales each error according to the size of the actual and hence needs no adjustment. The potential disadvantage of this method is that a few small loads ($a_k<<1$) which are poorly estimated could skew the average errors.

\section{Results}
\label{Sec:Results}

In this section we test and compare the methods we have developed in Section \ref{Methods}. The test period is the $53$ days consisting of $1^{st}$ October 2015 to $22^{nd}$ November 2015 inclusive. Since the temperature forecast data is available from hourly horizons up to 4 days ($96$ horizons) we consider up to four day ahead forecasts even when not considering the temperature variables. As described in the Methods section we will construct both point and probabilistic forecasts which will be evaluated using MAPE/RMAE and RCRPS respectively. The test period does not have any holiday dates but does contain the daylight savings date on the $25^{th}$ October. However this date will not be treated specially in this trial. Holiday dates will be considered in future work. 

\subsection{Average Errors}
\label{Sec:Errors}
To begin, we consider forecasting techniques without temperature inputs, this will be considered in much more detail in Section \ref{sec:Forecast Weather}. The average error scores for the methods are shown in Table \ref{AllErrorsAllMethods}, these consist of the errors for all four day ahead forecasts over the test period. First, comparing the benchmarks it is clear that the simple averages SMA are the best methods for point forecasts and in fact quite competitive with all methods. In fact it only has an average MAPE of $7\%$ worse than the best method, ARWDY. The LY method performs the worst and there is a slight improvement by using the yesterday-as-today estimate LD. Clearly the same period of the week forecast, LW, is the best of the random walk methods and shows the strong weekly periodicity of the data. 

The most accurate methods for point forecasts are ARWD, ARWDY and HWT which all include seasonal and autoregressive components. The ARWD and AWRDY methods are the best forecasts with the ARWD slightly better in terms of the RMAE. The HWT is the next best method and has a MAPE of only $1.3\%$ larger than the ARWD/ARWDY methods.  The ST and SnT methods is similar to the AWRDY method but without the autoregressive components and although improve on the benchmarks, only outperforms the SMA-4W by $2\%$. On average the KDE methods perform the worst only outperforming the random walk methods. Conditioning the KDE forecasts on the week period CKD-W improves the forecasts, especially with respect to the RMAE. 

For the probabilistic forecasts the empirical benchmark is also shown to be effective and gives a RCRPS only $20\%$ greater than the ARWD method. The ARWD is once again the best scoring forecast, closely followed by ARWDY. This time the ST forecast performs slightly better than the HWT forecast. Of the 100 feeders, the ARWD and ARWDY forecasts were the best performing, but not for all 100 feeders. Hence, there does not appear to be a one size fits all forecast which performs best for all feeders and identifying indicators of which forecast to choose will be an important step for practitioners.

\begin {table}[!htbp]
\begin{center}
  \begin{tabular}{ |c|c|c|c| }
\hline
 & \multicolumn{3}{c|}{Error Scores \%}\\
\hline 

Method &  MAPE (std)   &  RMAE (std) & RCRPS (std) \\
\hline
LD 												&  20.26 (5.43) & 20.31 (4.86) & -\\
LW                        &  18.67 (5.85) & 18.93 (5.34) & - \\
LY 												&  23.44 (7.12) & 23.12 (6.35) & - \\
SMA-4W                  	&  15.72 (4.98) & 16.53 (4.83) & - \\
SMA-5W                  	&	 15.73 (5.05) & 16.77 (5.03) & - \\
SMA		  									&  15.80 (5.04) & 17.02 (5.26) & - \\
Empirical 								&  16.19 (5.19) & 16.96 (4.95) & 12.62 (4.06) \\
\hline
HWT			 									&  14.84 (4.60) & 15.01 (4.14) & 11.06 (3.03) \\
ARWD 									   	& \textbf{14.65 (4.71)} & \textbf{14.67 (4.12)} & \textbf{10.30 (2.88)}  \\
ARWDY   									& \textbf{14.64 (4.55)} & \textbf{14.80 (4.15)} & \textbf{10.68 (2.91)} \\
ST                       	& 15.42 (5.20) & 15.42 (4.75) & 10.97 (3.31) \\
SnT                       & 15.66 (5.13) & 15.57 (4.70) & 11.10 (3.31) \\
KDE-W           					& 17.05 (5.56) & 19.36 (6.80) & 13.79 (4.52)   \\
KDE-W$\lambda$						& 17.08 (5.58) & 19.41 (6.83) & 13.80 (4.52) \\ 
CKD-W											& 16.54 (6.61) & 17.22 (6.25) & 13.23 (4.75) \\
\hline
  \end{tabular}
\end{center}
\caption{\label{AllErrorsAllMethods} MAPEs, RMAEs and RCRPSs for all forecast methods over all 4 day-ahead horizons for the entire 53 day test period. The lowest errors for each score are highlighted in bold. Standard deviations of the average errors between feeders are indicated in the brackets.} 
\end{table}

\begin{figure}
\begin{center}
\includegraphics[width=12cm]{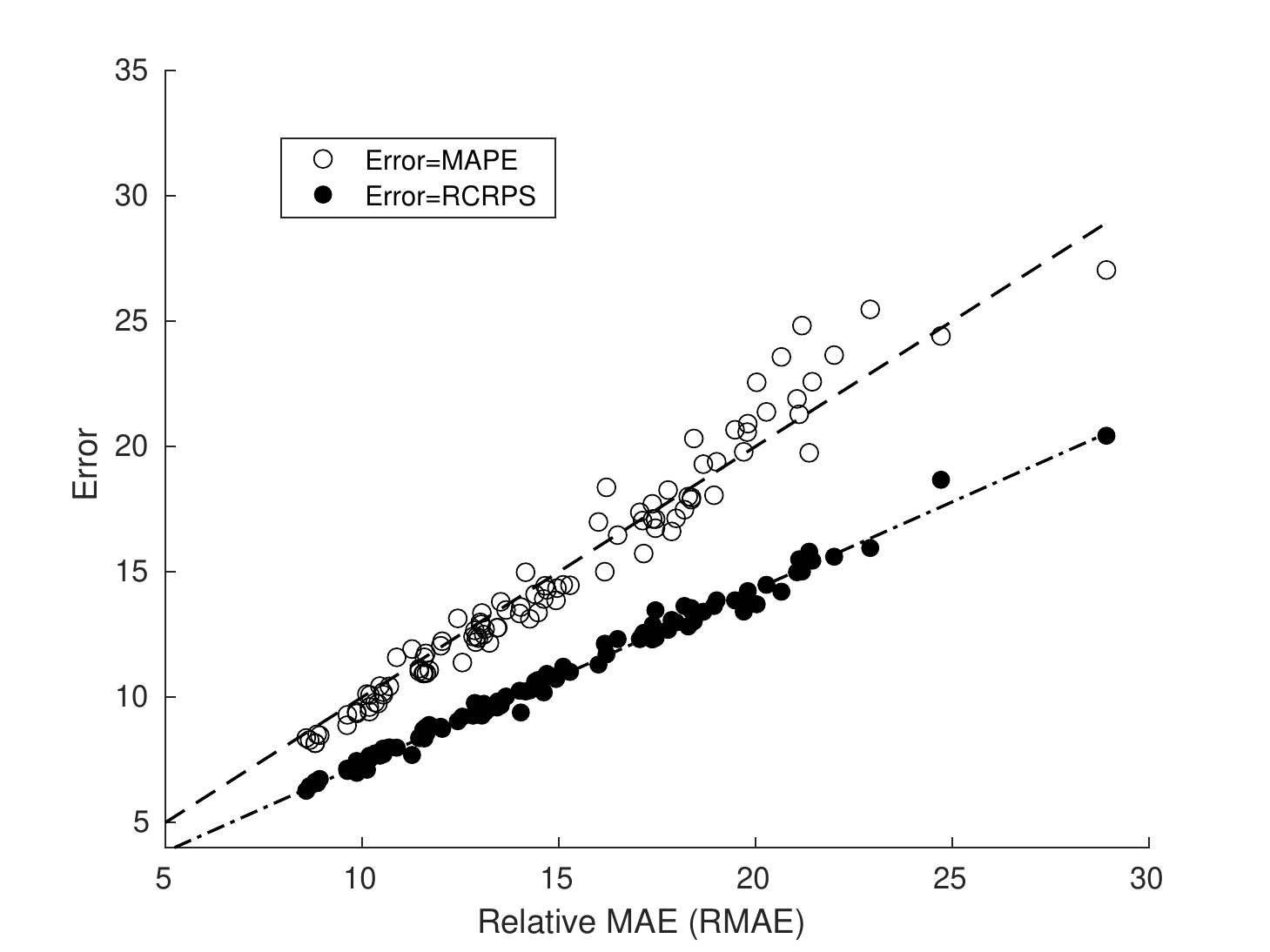}
\caption{\label{ComparisonErrorScoresARWDY} Scatter plot showing the average RCRPS (filled) and average MAPE (unfilled) versus average RMAE for each feeder. Also shown are lines of best fit. These results are for the ARWDY method.}
\end{center}
\end{figure}

If we compare the average errors for each feeder for a particular method we see that the errors are strongly correlated. Figure  \ref{ComparisonErrorScoresARWDY} shows this comparison for the ARWDY method. The plots are very similar for all methods. As expected, the RCRPS and RMAE are strongly related and this corresponds to a very strong linear correlation in the average errors (0.995). The MAPE and RMAE are also strongly correlated (0.981) but with more scatter, especially for larger errors. The strong correlation between the error measures means it is inefficient to present the remaining results in terms of all scores, MAPE, RMAE, and RCRPS. For this reason, and because of the ubiquitous use in the load forecasting community we will frame the rest of our discussion and analysis with respect to MAPE.

\subsection{Forecast Accuracy and Horizon}

Here we investigate the drop in forecast accuracy as a function of horizon. Initially we consider the forecast accuracy in terms of full days ahead. In other words we consider the accuracy of forecasts up to 1 day ahead, forecasts between 1 and 2 days, etc. The MAPEs as a function of whole days are shown in Table \ref{MAPErrorsSSPerDay} for selected methods. We note that the benchmarks forecast accuracy doe not change over the 4 day horizon since they are performed a week in advance. As expected, the most accurate forecasts horizon is one day ahead and the least is 4 days ahead. However, the drop in average accuracy is quite small, with no more than a $4\%$ drop in forecast score. 

\begin {table}[h]
\begin{center}
  \begin{tabular}{ |c|c|c|c|c| }
\hline
 & \multicolumn{4}{|c|}{MAPE}\\
 \hline
 Method & Day 1 & Day 2& Day3 & Day4 \\
\hline
HWT 						& 14.56 (4.46) & 14.83 (4.59) & 14.95 (4.67) & 15.04 (4.72) \\
ARWD     	 			& 14.42 (4.60) & 14.63 (4.68) & 14.74 (4.79) & 14.81 (4.80) \\
ARWDY    	 			& 14.34 (4.45) & 14.59 (4.52) & 14.75 (4.62) & 14.87 (4.65) \\
ST       	 			& 15.36 (5.15) & 15.41 (5.19) & 15.44 (5.22) & 15.49 (5.24) \\
SnT    		 			& 15.62 (5.11) & 15.65 (5.13) & 15.66 (5.14) & 15.70 (5.16) \\
KDE-W      			& 16.91 (5.45) & 17.00 (5.54) & 17.08 (5.60) & 17.21 (5.61) \\
KDE-W$\lambda$	& 16.94 (5.46) & 17.03 (5.55)	& 17.11	(5.62) & 17.25 (5.63) \\
CKD-W 					& 16.64 (6.67) & 16.56 (6.60) & 16.48 (6.57) & 16.50 (6.50) \\
\hline
  \end{tabular}
\end{center}
\caption{\label{MAPErrorsSSPerDay} MAPE Scores for each method over each day ahead horizon. Standard deviations of the average score across feeders are shown in the brackets.} 
\end{table} 

\begin{figure}
\begin{center}
\includegraphics[width=12cm]{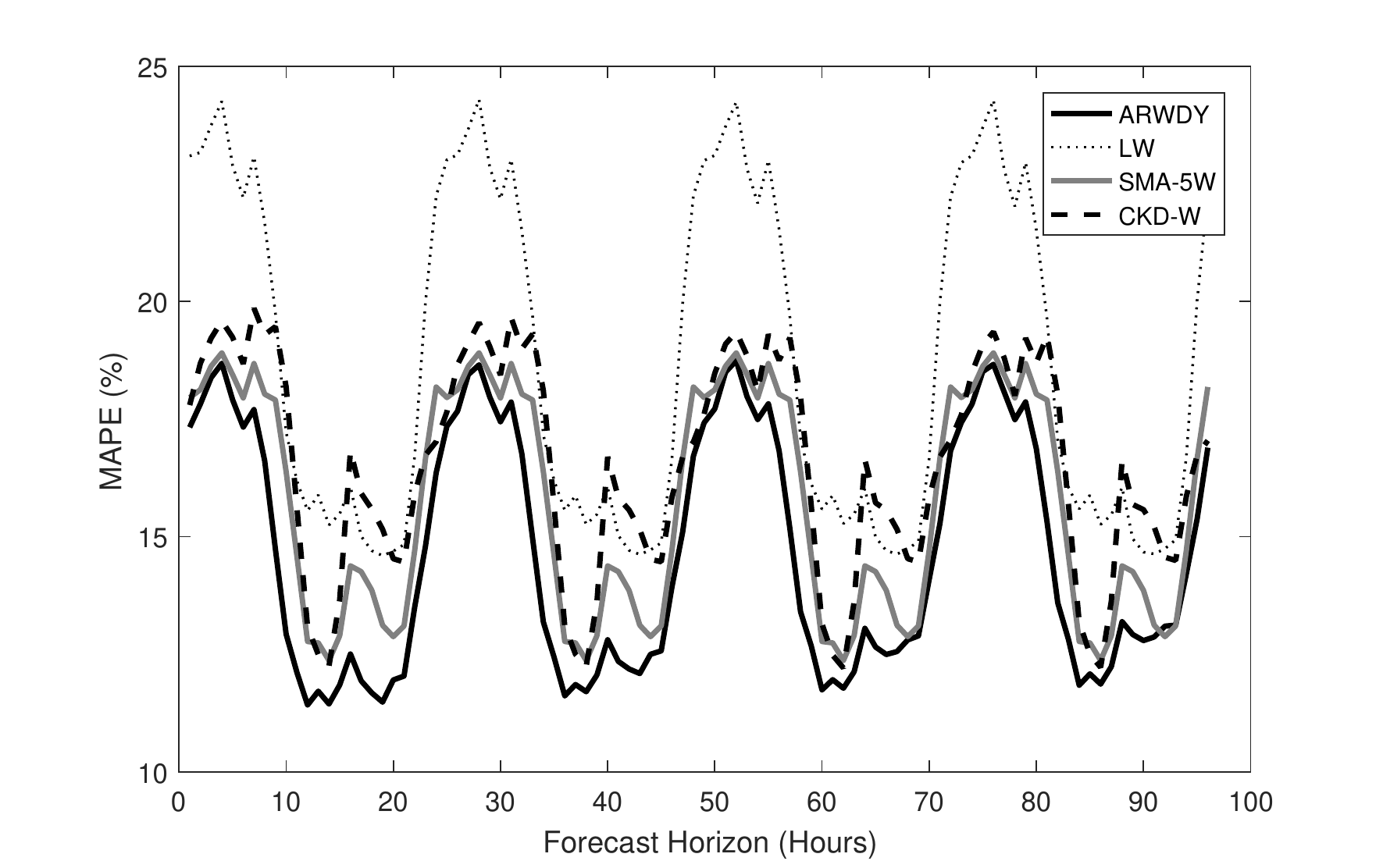}
\caption{\label{HorizonMAPEplotNew} Plot of average MAPE for selected methods for horizons from 1 hour to 96.}
\end{center}
\end{figure}

\begin{figure}
\begin{center}
\includegraphics[width=12cm]{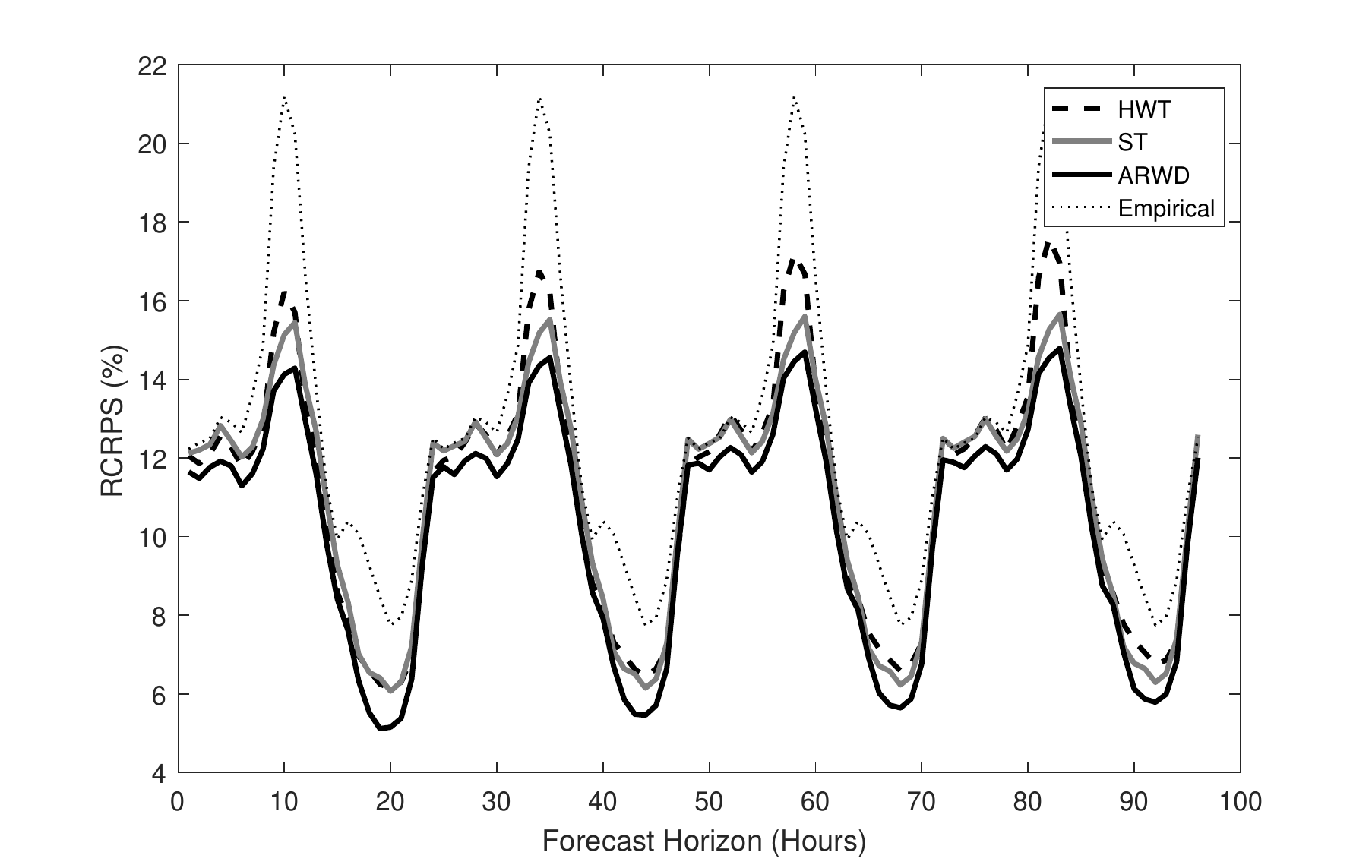}
\caption{\label{HorizonCRPS} Plot of average relative CRPS for selected methods for horizons from 1 hour to 96.}
\end{center}
\end{figure}

The MAPEs as a function of horizon at the hourly resolution for selected methods are shown in Figure \ref{HorizonMAPEplotNew}. We note that the results are similar whatever the methods considered. First recall that the first horizon corresponds to the period $8-9$AM. There are a number of interesting observations. It is clear that all forecast methods produce a similar shape and the more accurate forecasts (ARWDY for example) have smaller errors at all horizons than the least accurate forecasts. Secondly, as confirmed with Table \ref{MAPErrorsSSPerDay} there is a general trend with a small overall increase in the error as a function of horizon. However, the strongest driver of the forecast accuracy is the period of the day. The most inaccurately forecast time periods corresponds to the hours from around 6AM to 6PM. Similarly the period most easily estimated is the evening and night period. Surprisingly the evening period (from 6PM until 11PM) we would expect to be quite difficult to forecast. In fact, what we discover is that the horizon-error shape may be an artifact of the error measure used. In Figure \ref{HorizonCRPS} we show a similar plot but this time for the relative CRPS score for selected methods. This shows the expected larger error in the evening period. The difference between the two scoring functions is that the MAPE (\ref{MAPE}) normalises each hourly forecast error with respect to the demand at the same hour. Since for residential feeders (which dominate the composition of the feeders in this trial) have the largest demand in this evening period the MAPEs are smaller compared to the RCRPS or RMAE. This could have important implications for choosing between different error measures for LV forecasts, especially for the many applications where peak demand is of the most importance, such as in peak demand reduction via storage devices \cite{Rowe2014b}. It should be noted that the horizon plot shape shown in Figure \ref{HorizonCRPS} is consistent with that as shown by the authors in \cite{Arora2016} for 800 individual residential customers. 

\subsection{Accuracy and Feeder Size}

As described in the data analysis section the connectivity of the feeders varies with different mixtures (domestic and non-domestic) and numbers of customers.  Recent literature has shown there is a link between size of the aggregation (which makes up a demand time series) and the accuracy of the forecast \cite{Humeau2013, Sevlian2014}. In contrast to the previous work, we study real LV networks and hence may have slightly different behaviour compared to aggregations of smart meter data. These networks includes street furniture such as street-lighting and traffic lights, and also means there may be similarities in consumption between the different consumers on the network which is not captured by randomly selected smart meters.

\begin{figure}
\begin{center}
\includegraphics[width=12cm]{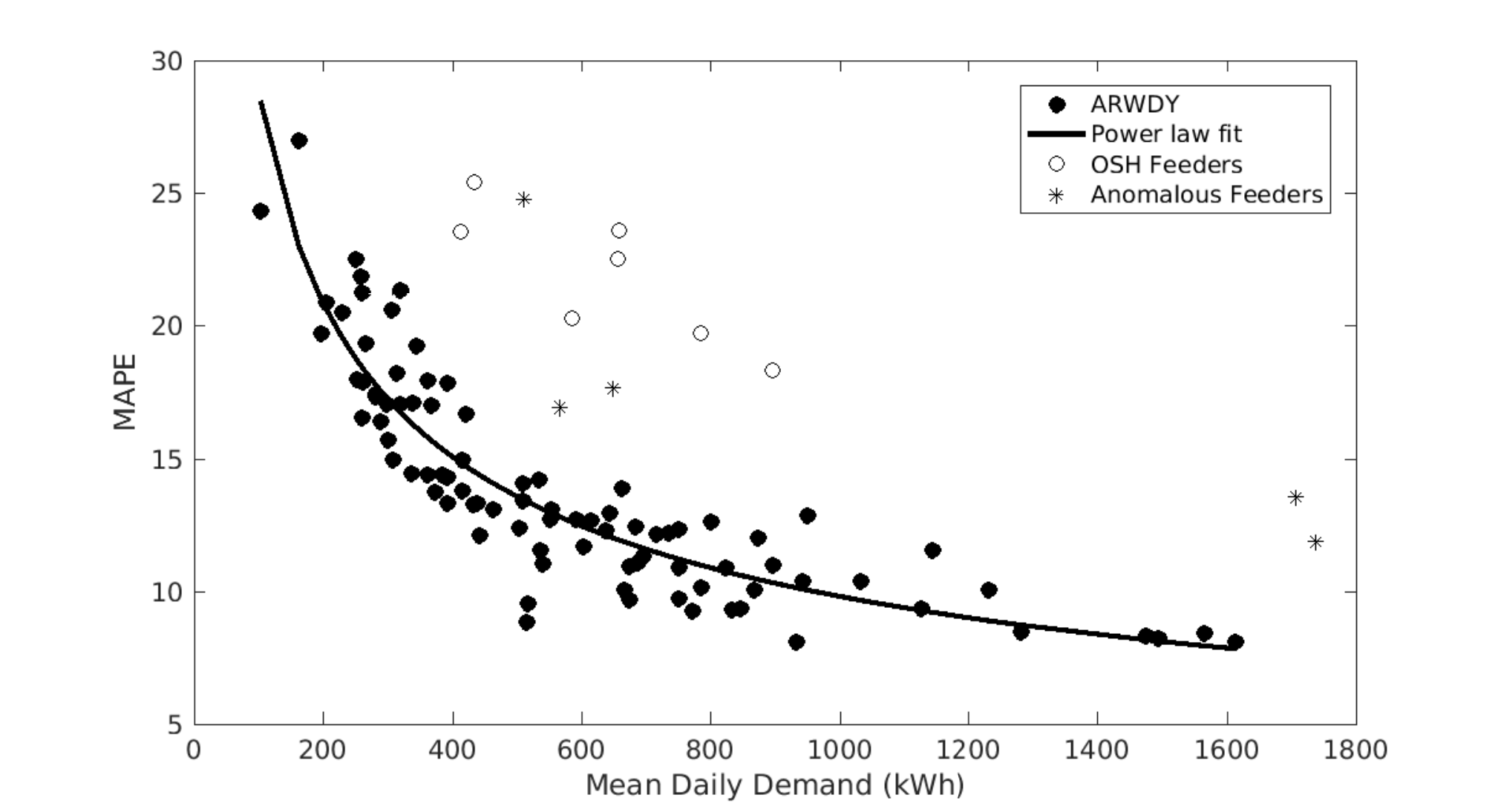}
\caption{\label{ScaleLawNew} Scatter plot of the relationship between MAPE and mean daily demand for two different forecasting methods. Feeders which apparently have overnight storage heaters and those with unexplained larger errors have been labeled separately as OSH and anomalous respectively. Also shown is a power law fit to the non-OSH/anomalous feeders.}
\end{center}
\end{figure}

Figure \ref{ScaleLawNew} shows the MAPEs for each individual feeder as a function of the average daily demand. The majority of the feeders appear to fit a power law relationship which is consistent with the results found in \cite{Humeau2013, Sevlian2014}. It was clear that twelve of the feeders did not fit the relationship as tightly as the remaining 88 feeders. On closer inspection seven of these feeders had very large overnight demands and were likely the result of overnight storage heaters (OSH). Feeders with a high proportion of OSH will have a high demand (especially in the winter) for relatively low numbers of customers. However, despite their higher daily mean demand they will have the volatility of a standard feeder but of smaller size. The other four anomalous feeders could not be characterised as easily, although the two largest were found to be likely dominated by commercial consumers due to the large demand during the daytime period. The cause of the errors was not obvious. We fit a power law curve to the 88 other feeders as shown in Figure \ref{ScaleLawNew}. If the customers producing the aggregated demand were independent and identically distributed (IID) we would expect an exponent of $-0.5$ however we found an exponent equal to $-0.47$ indicating the IID assumption is not completely accurate. Further we found the the variation of the customers demand also followed a power law curve very similar to the mean errors (not shown).

\subsection{Weather Effect Analysis}
\label{sec:Forecast Weather}

Weather, in particular those related to temperature, often plays an important role in the accuracy of a load forecasts for high voltage level substations \cite{Hobby2011, Hong2016}. In this section we consider in more detail the impact of including temperature in the forecasts. In particular, we consider both ex-ante and ex-post forecasts by utilising either forecast or actual temperature values respectively. In reality, ex-ante are the practical way to create true forecasts since, the actual temperature data is not available ahead of time. However, we include the ex-post forecasts here for comparison since much of the literature is based on these forms of forecast. Table \ref{ErrorsNonWeather_NewMAPE} shows the MAPEs for the average 4 day ahead forecasts over the test period for selected methods including their updates using temperature data, both actual and forecast values as input. From the table it is clear that the inclusion of temperature (either actual or forecast) has minimal effect on the forecast accuracy. In fact for ARWD, ARWDY and CKD-W, including the temperature is detrimental to the forecast accuracy. For the ST and SnT method, there are inconsistent results, using the actual temperature values has little to no affect on the forecast accuracy. Using the forecast values has a small affect, with at most a $1.7\%$ reduction in forecast accuracy.

\begin {table}[h]
\begin{center}
  \begin{tabular}{ |c|c|c|c| }
\hline
 &\multicolumn{3}{c|}{Temperature Type}\\
 \hline
Method &  None   & Forecast & Actual \\
\hline
ARWD  					&  14.65 (4.71) & 20.17 (7.52) &  20.03 (7.48) \\
ARWDY 					&  14.64 (4.55) & 15.36 (4.60) &  15.16 (4.56) \\
ST              &  15.42 (5.20) & 15.16 (4.92) &  15.39 (5.11) \\
SnT             &  15.66 (5.13) & 15.48 (4.92) &  15.66 (5.07) \\
CKD-W					  &  16.54 (6.61) & 17.16 (7.03) &  17.02 (7.02) \\
\hline
  \end{tabular}
\end{center}
\caption{\label{ErrorsNonWeather_NewMAPE} MAPEs for the methods showing the effect of including temperature data (actual or forecast) for a selection of methods. Standard deviations of the average MAPEs across the feeders are shown in brackets.} 
\end{table}

Further, Table \ref{MAPErrorsSSPerDay2} shows the accuracy of the ex-ante forecasts as a function of day ahead horizons. Also included is the MAPEs of the temperature forecasts themselves. The ARWD and ARWDY forecasts drop in accuracy by $5.6\%$ and $4.3\%$ respectively (from 1 to 4 day ahead) whereas the ST and SnT forecasts hardly change in accuracy at all. The CKD-W forecast actually improves at the 3 day ahead horizon compared to the 1 day and 2 day ahead forecasts. These results are in contrast to the accuracy of the temperature forecasts themselves, which drop in accuracy by more than $80\%$ from one day ahead to four days ahead. If the weather was a major driver for the load we would expect a much larger drop in accuracy with horizon. Further, we also considered including up to two lags of the temperature data within the forecast methods to see if they had any but this also had no effect on the accuracy of the forecasts either. We note that the CKD-W method naturally contains lags within the model. Although the short lags of temperature enter the CKD-WT method through the kernel of the temperature, the inclusion of the temperature in CKD-W results in less accurate forecasts, indicating that short lags do not contribute to more accurate forecasts.

\begin {table}[h]
\begin{center}
  \begin{tabular}{ |c|c|c|c|c| }
\hline
 & \multicolumn{4}{|c|}{MAPE}\\
 \hline
 Method & Day 1 & Day 2& Day3 & Day4 \\
\hline
ARWD 			& 19.52 (7.10) & 20.10 (7.54) &  20.46 (7.69) & 20.61 (7.76) \\
ARWDY 		& 15.00 (4.54) & 15.30 (4.56) &  15.50 (4.68) & 15.65 (4.67) \\
ST 		  	& 15.12 (4.91) & 15.21 (4.94) &  15.16 (4.94) & 15.16 (4.91) \\
SnT   		& 15.47 (4.92) & 15.53 (4.95) &  15.46 (4.93) & 15.47 (4.92) \\
CKD-WTf		& 17.22 (7.11) & 17.24 (7.10) &  17.02 (6.96) & 17.19 (6.88) \\
\hline
\hline
Temperature &  8.98 & 10.57 & 13.46 & 16.47 \\
\hline
  \end{tabular}
\end{center}
\caption{\label{MAPErrorsSSPerDay2} MAPE Scores for different day ahead horizons for a selection of methods based on utilising forecast temperature values. Standard deviations of the average feeder errors are presented in the brackets. Also for comparison is the average MAPE for the temperature forecast themselves.} 
\end{table} 

We investigate the temperature effect in more detail for the ARWDY forecast which is one of the better methods and has only a small drop in accuracy by the inclusion of the temperature. We only consider day ahead forecasts since this is when the temperature forecasts are most accurate (as shown in Table \ref{MAPErrorsSSPerDay2}). First we observe, from comparing the MAPEs for each feeder, that including the temperature forecasts only improves the errors for 7 of the 100 feeders. If we use the temperature actuals in the forecasts this only improves the forecasts for 10 feeders. In all cases the MAPEs do not improve by any more than $1.5\%$.

To further test the effect of including the temperature we also consider comparing the change in the distribution of the errors. We do this by using the two-sample Kolmogorov-Smirnov test (implemented using \textit{kstest2} in Matlab) at the $5\%$ significance level. Since we have observed that the different hours of the day have different distributions of demand we split the errors according to both hour ($1$ to $24$) and feeder. This gives us $2400$ distributions to compare. Comparing the ARWDY without and with the temperature forecasts as input we find that the null hypothesis (of the errors coming from the same distribution) is only rejected for a total of 40 distributions. These are the result of 24 feeders, with no more than 4 distributions failing the null hypothesis on any one of these feeders. Further to this, of the 40 distributions failing the null hypothesis $35$ occur between midnight and 6AM. In other words, utilising temperature in the forecasts, around 87$\%$ of significant changes to the forecast accuracies occur during the early morning hours when demand is usually more stable. In addition to this when the distributions are significantly changed by including the weather, the accuracy is only improved for $16$ of the $40$ distributions. Similar results hold when using the actual temperature but only $33$ distributions are significantly changed (i.e. reject the null hypothesis). 

\begin{figure}
\begin{center}
\includegraphics[width=10cm]{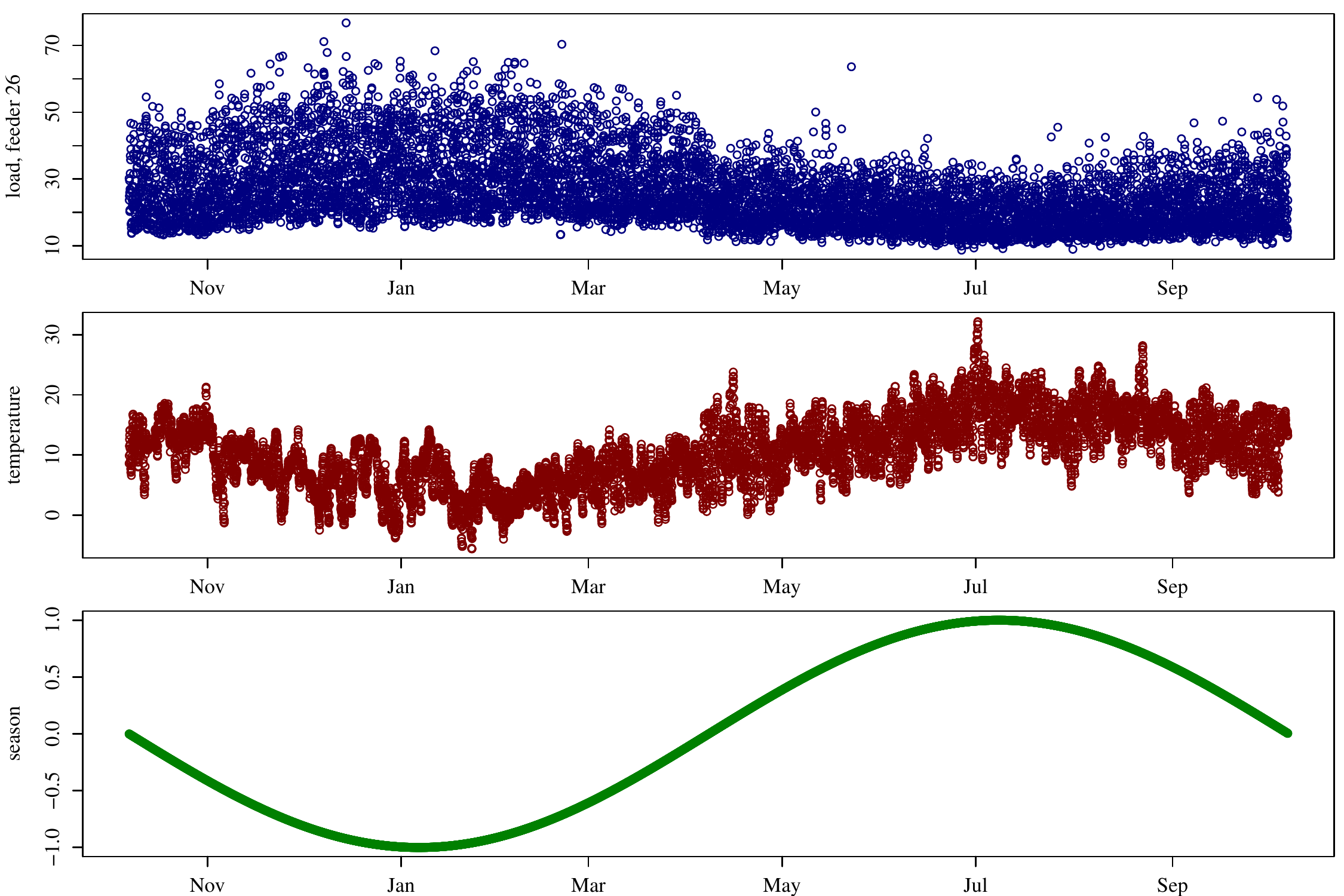}
\caption{\label{3ts_feeder1026} Load curve (top) for feeder 26 with yearly temperature profile (centre) and a simple seasonal profile (bottom).}
\end{center}
\end{figure}

As shown in Section \ref{Sec:DataAnal}, for some feeders there is a strong correlation between the load and the temperature. However, as we have seen in this section, when we include the effect in our model the accuracy either changes only slightly or, in the case of ARWD, gets significantly worse. A major difference between the ARWD methods and the other methods is the lack of a seasonal term. In fact, demand is much more strongly correlated with seasonality then temperature. If seasonality is a stronger driver of demand than temperature then this could result in the detrimental performance in ARWD when included in the model. The strong relationship between demand, temperature and seasonality (represented by a simple sinusoidal curve) is illustrated in Figure \ref{3ts_feeder1026}. By treating the temperature as a surrogate for the seasonality the ARWD model may erroneously over-train on data which is not related to the load.

The evidence thus suggests that at least in this area of the UK there is not a strong causal link between demand and temperature. Seasonality is a stronger driver of the demand. However, we are careful to extrapolate this to other areas since we have only considered a relatively small area of the UK and only two months of the winter period in 2015. Further, more research is required on both geographical and culture comparisons to make firmer conclusions.

\section{Discussion}
\label{Sec:Discussion}

Short term load forecasts at the low voltage (LV) level are becoming increasingly important as electricity networks prepare for a low carbon future. Network solutions such as storage devices and energy management systems will require accurate forecasts to optimise the headroom and potential cost savings. Although there is a large literature in short term load forecasting techniques there is not much investigation or results for LV level demand. Such demand is much more volatile and challenging then high voltage systems and there is still much to learn about the best methods and inputs for accurate forecasts. 

In this paper we have presented a number of short term forecasting methods, both point and probabilistic, and tested them on 100 real LV feeders. We also compared them to several benchmarks, some of which are quite competitive. As a result of the studies we have found some interesting results. Firstly, accurate results can be obtained by relatively simple methods, in particular a simple average of the previous four weeks performed quite well. The best performing method were those based on autoregression methods and a Holt-Winters-Taylor exponential smoothing method. We also illustrated some important drivers for the accuracy of the forecast. Firstly, the size of the feeder was a major determining factor for forecast accuracy with smaller feeders much more difficult to predict than larger feeders. This could have important implications for network planning decisions, e.g. identifying the optimal investments and location for storage devices. Secondly, we found that the time period of the day was a major indicator of the accuracy and was more important than the forecast horizon. 

In contrast to HV level load forecasts, temperature was not an important factor in the accuracy of our forecasts. We presented some detailed analysis of the results to show that often the temperature either had little-to-no effect on the forecast accuracy. However in many cases including temperature was actually detrimental to the accuracy. One potential explanation for this is the strong correlation between temperature and seasonality. By erroneously training on the temperature as a surrogate for seasonality may produce large errors than not including temperature at all. 

Finally, we performed some empirical comparison of the forecasts using a variety of error measures. It was found that there was a strong correlation between the scores. Hence the MAPE/MAE scores on the point wise version of the forecasts could be used to provide an accurate indication of the accuracy of the corresponding probabilistic forecast. This supports work recently in \cite{Xie2016} and could have implications for reducing the cost of model selection for probabilistic load forecasts.

The research presented here can support better understanding of low voltage short term load forecasts and can form the foundation for deeper insights and further work. In particular there is still further understanding into the effects of weather, the role of LV network connectivity and better probabilistic forecasts.

\section*{Acknowledgements}
We would like to acknowledge the support of Scottish and Southern Electricity Networks (SSEN) and for the funding of this project via the Low Carbon Network Fund Project the New Thames Valley Vision Project (SSET203 New Thames Valley Vision). We would also like to thank Dr Tamsin Lee for her support and early discussions on the research. 

\bibliographystyle{plain}
\bibliography{bibfile}

\begin{thebibliography}{10}

\bibitem{Alberg2017}
Dima Alberg and Mark Last.
\newblock {\em Short-Term Load Forecasting in Smart Meters with Sliding
  Window-Based ARIMA Algorithms}, pages 299--307.
\newblock Springer International Publishing, Cham, 2017.

\bibitem{Alfares2002}
H.~Alfares and M.~Nazeeruddin.
\newblock Electric load forecasting: literature survey and classification of
  methods.
\newblock {\em International Journal of Systems Science}, 33:23--34, 2002.

\bibitem{Alzate2013}
C.~Alzate and M.~Sinn.
\newblock Improved electricity load forecasting via kernel spectral clustering
  of smart meters.
\newblock In {\em 2013 IEEE 13th International Conference on Data Mining},
  pages 943--948, 2013.

\bibitem{Arora2016}
S.~Arora and J.~Taylor.
\newblock Forecasting electricity smart meter data using conditional kernel
  density estimation.
\newblock {\em Omega}, 59:47--59, 2016.

\bibitem{Arora2017}
S.~Arora and J.~Taylor.
\newblock Rule-based autoregressive moving average models for forecasting load
  on special days: A case study for france.
\newblock {\em European Journal of Operational Research}, 266:259--268, 2017.

\bibitem{Bennett2014}
Christopher Bennett, Rodney~A. Stewart, and Junwei Lu.
\newblock Autoregressive with exogenous variables and neural network short-term
  load forecast models for residential low voltage distribution networks.
\newblock {\em Energies}, 7(5):2938--2960, 2014.

\bibitem{Bennett2014b}
Christopher~J. Bennett, Rodney~A. Stewart, and Jun~Wei Lu.
\newblock Forecasting low voltage distribution network demand profiles using a
  pattern recognition based expert system.
\newblock {\em Energy}, 67(Supplement C):200 -- 212, 2014.

\bibitem{Bessec2008}
Marie Bessec and Julien Fouquau.
\newblock The non-linear link between electricity consumption and temperature
  in europe: {A} threshold panel approach.
\newblock {\em Energy Economics}, 30(5):2705 -- 2721, 2008.

\bibitem{Chitsaz2015}
Hamed Chitsaz, Hamid Shaker, Hamidreza Zareipour, David Wood, and Nima Amjady.
\newblock Short-term electricity load forecasting of buildings in microgrids.
\newblock {\em Energy and Buildings}, 99(Supplement C):50 -- 60, 2015.

\bibitem{Dahl2017}
Magnus Dahl, Adam Brun, and Gorm~B. Andresen.
\newblock Using ensemble weather predictions in district heating operation and
  load forecasting.
\newblock {\em Applied Energy}, 193:455 -- 465, 2017.

\bibitem{Dangha2017}
T.-H. {Dang-Ha}, F.~M. {Bianchi}, and R.~{Olsson}.
\newblock {Local Short Term Electricity Load Forecasting: Automatic
  Approaches}.
\newblock {\em ArXiv e-prints}, 2017.

\bibitem{Dehalwar2016}
V.~Dehalwar, A.~Kalam, M.~L. Kolhe, and A.~Zayegh.
\newblock Electricity load forecasting for urban area using weather forecast
  information.
\newblock In {\em 2016 IEEE International Conference on Power and Renewable
  Energy (ICPRE)}, pages 355--359, 2016.

\bibitem{Ding2016}
N.~Ding, C.~Benoit, G.~Foggia, Y.~Bésanger, and F.~Wurtz.
\newblock Neural network-based model design for short-term load forecast in
  distribution systems.
\newblock {\em IEEE Transactions on Power Systems}, 31(1):72--81, 2016.

\bibitem{Elbaz2015}
Wessam El-Baz and Peter Tzscheutschler.
\newblock Short-term smart learning electrical load prediction algorithm for
  home energy management systems.
\newblock {\em Applied Energy}, 147:10 -- 19, 2015.

\bibitem{Gerossier2017}
Alexis Gerossier, Robin Girard, Georges Kariniotakis, and Andrea Michiorri.
\newblock Probabilistic day-ahead forecasting of household electricity demand.
\newblock In {\em Proceedings of the 24th International Conference on
  Electricity Distribution (CIRED)}, 2017.

\bibitem{Ghofrani2011}
Mahmoud Ghofrani, Mohammad Hassanzadeh, Mehdi Etezadi-Amoli, and M~Sami Fadali.
\newblock Smart meter based short-term load forecasting for residential
  customers.
\newblock In {\em IEEE North American Power Symposium (NAPS)}, pages 1--5,
  Boston, USA, 2011.

\bibitem{Giasemidis2017}
Georgios Giasemidis, Stephen Haben, Tamsin Lee, Colin Singleton, and Peter
  Grindrod.
\newblock {A genetic algorithm approach for modelling low voltage network
  demands}.
\newblock {\em Applied Energy}, 203:463--473, 2017.

\bibitem{Gneiting2007}
Tilmann Gneiting and Adrian~E Raftery.
\newblock Strictly proper scoring rules, prediction, and estimation.
\newblock {\em Journal of the American Statistical Association}, 102:359--378,
  2007.

\bibitem{Haben2016b}
S.~Haben and G.~Giasemidis.
\newblock A hybrid model of kernel density estimation and quantile regression
  for gefcom2014 probabilistic load forecasting.
\newblock {\em Int. J. Forecasting}, 32:1017--1022, 2016.

\bibitem{Haben2014}
S.~Haben, J.~A. Ward, D.~V. Greetham, P.~Grindrod, and C.~Singleton.
\newblock A new error measure for forecasts of household-level, high resolution
  electrical energy consumption.
\newblock {\em Int. J. of Forecasting}, 30:246--256, 2014.

\bibitem{Hayes2015}
B.~Hayes, J.~Gruber, and M.~Prodanovic.
\newblock Short-term load forecasting at the local level using smart meter
  data.
\newblock In {\em 2015 IEEE Eindhoven PowerTech}, pages 1--6, 2015.

\bibitem{Hobby2011}
John~D. Hobby and Gabriel~H. Tucci.
\newblock Analysis of the residential, commercial and industrial electricity
  consumption.
\newblock In {\em Innovative Smart Grid Technologies Asia (ISGT), 2011 IEEE
  PES}, Perth, Australia, 2011.

\bibitem{Hong2014}
T.~Hong.
\newblock Long term probabilistic load forecasting and normalization with
  hourly information.
\newblock {\em IEEE Transactions on Smart Grid}, 5:456--462, 2014.

\bibitem{Hong2016}
T.~Hong, P.~Pinson, S.~Fan, H.~Zareipour, A.~Troccoli, and R.~J Hyndman.
\newblock Probabilistic energy forecasting: Global energy forecasting
  competition 2014 and beyond.
\newblock {\em International Journal of Forecasting}, 32:896--913, 2016.

\bibitem{Hong2016b}
Tao Hong and Shu Fan.
\newblock Probabilistic electric load forecasting: A tutorial review.
\newblock {\em International Journal of Forecasting}, 32(3):914 -- 938, 2016.

\bibitem{HONG2014357}
Tao Hong, Pierre Pinson, and Shu Fan.
\newblock Global energy forecasting competition 2012.
\newblock {\em International Journal of Forecasting}, 30(2):357 -- 363, 2014.

\bibitem{Humeau2013}
S.~Humeau, T.~K. Wijaya, M.~Vasirani, and K.~Aberer.
\newblock Electricity load forecasting for residential customers: Exploiting
  aggregation and correlation between households.
\newblock In {\em 2013 Sustainable Internet and ICT for Sustainability
  (SustainIT)}, pages 1--6, Oct 2013.

\bibitem{Irish2012}
{{Irish} Social Science Data Archive}.
\newblock Cer smart metering project, 2012.

\bibitem{Jeon2012}
Jooyoung Jeon and James~W. Taylor.
\newblock Using conditional kernel density estimation for wind power density
  forecasting.
\newblock {\em Journal of the American Statistical Association},
  107(497):66--79, 2012.

\bibitem{Koenker1978}
R.~Koenker and G.~Bassett Jr.
\newblock Regression quantiles.
\newblock {\em Econometrica}, 46:33--50, 1978.

\bibitem{Li2017}
Z.~Li, A.S. Hurn, and A.E. Clements.
\newblock Forecasting quantiles of day-ahead electricity load.
\newblock {\em Energy Economics}, 67:60 -- 71, 2017.

\bibitem{Liu2017}
Y.~Liu, W.~Wang, and N.~Ghadimi.
\newblock Electricity load forecasting by an improved forecast engine for
  building level consumers.
\newblock {\em Energy}, 139:18--30--3052, 2017.

\bibitem{Quilumba2015}
F.~L. Quilumba, W.~J. Lee, H.~Huang, D.~Y. Wang, and R.~L. Szabados.
\newblock Using smart meter data to improve the accuracy of intraday load
  forecasting considering customer behavior similarities.
\newblock {\em IEEE Transactions on Smart Grid}, 6(2):911--918, 2015.

\bibitem{Rahman1990}
S.~Rahman.
\newblock Formulation and analysis of a rule-based short-term load forecasting
  algorithm.
\newblock In {\em Proceedings of the IEEE}, volume~78, pages 805--816, 1990.

\bibitem{Rowe2014b}
M.~Rowe, T.~Yunusov, S.~Haben, C.~Singleton, W.~Holderbaum, and B.~Potter.
\newblock A peak reduction scheduling algorithm for storage devices on the low
  voltage network.
\newblock {\em IEEE Trans. on Smart Grid}, 5:2115--2124, 2014.

\bibitem{Rowe2014a}
Matthew Rowe, Timur Yunusov, Stephen Haben, William Holderbaum, and Ben Potter.
\newblock {The real-time optimisation of DNO owned storage devices on the LV
  network for peak reduction}.
\newblock {\em Energies}, 7(6):3537--3560, 2014.

\bibitem{Sevlian2014}
R.~{Sevlian} and R.~{Rajagopal}.
\newblock {Short Term Electricity Load Forecasting on Varying Levels of
  Aggregation}.
\newblock {\em ArXiv e-prints}, March 2014.

\bibitem{Sousa2009}
J.~M.~C. Sousa, L.~M.~P. Neves, and H.~M.~M. Jorge.
\newblock Short-term load forecasting using information obtained from low
  voltage load profiles.
\newblock In {\em 2009 International Conference on Power Engineering, Energy
  and Electrical Drives}, pages 655--660, March 2009.

\bibitem{Sun2016}
X.~Sun, P.~B. Luh, K.~W. Cheung, W.~Guan, L.~D. Michel, S.~S. Venkata, and
  M.~T. Miller.
\newblock An efficient approach to short-term load forecasting at the
  distribution level.
\newblock {\em IEEE Transactions on Power Systems}, 31(4):2526--2537, 2016.

\bibitem{Taieb2016}
S.~Ben Taieb, R.~Huser, R.~J. Hyndman, and M.~G. Genton.
\newblock Forecasting uncertainty in electricity smart meter data by boosting
  additive quantile regression.
\newblock {\em IEEE Transactions on Smart Grid}, 7(5):2448--2455, 2016.

\bibitem{Taieb2017}
Souhaib~Ben Taieb, James~W. Taylor, and Rob~J. Hyndman.
\newblock Hierarchical probabilistic forecasting of electricity demand with
  smart meter data.
\newblock pages 1--30, 2017.

\bibitem{Taylor2003}
J.~W. Taylor.
\newblock Short-term electricity demand forecasting using double seasonal
  exponential smoothing.
\newblock {\em The Journal of the Operational Research Society}, 54:799--805,
  2003.

\bibitem{Taylor2008}
J.~W. Taylor and A.~Espasa.
\newblock Energy forecasting.
\newblock {\em International Journal of Forecasting}, 24:561–--565, 2008.

\bibitem{Valgaev2016}
O.~Valgaev, F.~Kupzog, and H.~Schmeck.
\newblock Low-voltage power demand forecasting using k-nearest neighbors
  approach.
\newblock In {\em 2016 IEEE Innovative Smart Grid Technologies - Asia
  (ISGT-Asia)}, pages 1019--1024, 2016.

\bibitem{Veit2014}
Andreas Veit, Christoph Goebel, Rohit Tidke, Christoph Doblander, and Hans-Arno
  Jacobsen.
\newblock Household electricity demand forecasting: Benchmarking
  state-of-the-art methods.
\newblock In {\em Proceedings of the 5th International Conference on Future
  Energy Systems}, e-Energy '14, pages 233--234, New York, NY, USA, 2014. ACM.

\bibitem{Xie2016}
J.~Xie and T.~Hong.
\newblock Comparing two model selection frameworks for probabilistic load
  forecasting.
\newblock In {\em 2016 International Conference on Probabilistic Methods
  Applied to Power Systems (PMAPS)}, pages 1--5, Oct 2016.

\bibitem{Yu2017}
C.~N. Yu, P.~Mirowski, and T.~K. Ho.
\newblock A sparse coding approach to household electricity demand forecasting
  in smart grids.
\newblock {\em IEEE Transactions on Smart Grid}, 8(2):738--748, 2017.

\bibitem{Zufferey2017}
Thierry Zufferey, Andreas Ulbig, Stephan Koch, and Gabriela Hug.
\newblock {\em Forecasting of Smart Meter Time Series Based on Neural
  Networks}, pages 10--21.
\newblock Springer International Publishing, Cham, 2017.

\end{thebibliography}

\end{document}